
%
%
\parindent=0truept
\def\ave#1{\langle#1\rangle}           
\rightline{TITCMT-95-22}
\bigskip
{}~~

\centerline{\bf Critical properties of the transition between the
Haldane phase and}
\medskip
\centerline{\bf the large-$D$ phase of the spin-1/2
ferromagnetic-antiferromagnetic}
\medskip
\centerline{\bf Heisenberg chain with on-site anisotropy}

\bigskip
\medskip
\centerline{Kiyomi Okamoto${}^\dagger$}
\medskip
\medskip
\centerline{Department of Physics, Tokyo Institute of Technology,}
\centerline{Oh-okayama, Meguro-ku, Tokyo 152, Japan}
\medskip
\bigskip
\rm
\centerline{(Received ~~~~~~~~~~~~~~~~~~~~~~~~~~~~1995)}
\bigskip
\bigskip
\bigskip
\bigskip
\hyphenation{an-i-so-tro-py}
{\bf Abstract.~~} We analytically study the ground-state quantum phase
transition between the Haldane phase and the large-$D$ (LD) phase of the
$S=1/2$ ferromagnetic-antiferromagnetic alternating Heisenberg chain
with on-site anisotropy.
We transform this model into a generalized version of the alternating
antiferromagnetic Heisenberg model with anisotropy.
In the transformed model, the competition between the transverse and
longitudinal bond alternations yields the Haldane-LD transition.
Using the bosonization method, we show that the critical exponents vary
continuously on the Haldane-LD boundary.
Our scaling relations between critical exponents very well explains the
numerical results by Hida.

\bigskip
\bigskip


\bigskip
\bigskip
\hrule
\bigskip
\bigskip
$\dagger$ E-mail address: kokamoto@stat.phys.titech.ac.jp

\vfill\eject

{\bf\S 1. Introduction}

\medskip

{}~~~~~Since Haldane's prediction[1,2], the integer spin chains have been
attracting much attention.
Hida [3,4] tried to elucidate the properties of the spin-1 Heisenberg
chain from the standpoint of the spin-1/2
ferromagnetic-antiferromagnetic alternating chain.
He considered the Hamiltonian [4]
$$
    {\cal H}_D
    = 2J \sum_{j=1}^N {\bf S}_{2j} \cdot {\bf S}_{2j+1}
      + 2J' \sum_{j=1}^N {\bf S}_{2j-1} \cdot {\bf S}_{2j}
      + D \sum_{j=1}^N (S_{2j-1}^z + S_{2j}^z)^2~,
    \eqno(1.1)
$$
where ${\bf S}_j$ is the spin-1/2 operator.
We assume that $J>0$ (antiferromagnetic) and $J'<0$ (ferromagnetic) and
we are interested in only the ground-state.
Hereafter we call the model (1.1) \lq\lq $D$-model\rq\rq \ following
Hida [5].

{}~~~~~In the case of $J' \rightarrow - \infty$, where the spins ${\bf
S}_{2j-1}$ and ${\bf S}_{2j}$ form a triplet, the model (1.1) reduces to
a spin-1 antiferromagnetic Heisenberg chain with on-site anisotropy:
$$
    \hat{\cal H}_D^{S=1}
    = {J \over 2} \sum_{j=1}^N \hat{\bf S}_j \cdot \hat{\bf S}_{j+1}
      + D \sum_{j=1}^N ({\hat S}_j^z)^2~,
    \eqno(1.2)
$$
where $\hat{\bf S}_j = {\bf S}_{2j-1} + {\bf S}_{2j}$ is the spin-1
operator.
When $J'=D=0$, on the other hand, Hamiltonian (1.1) reduces to the
two-spin problem.
Its ground-state is the complete dimer state in which the spins ${\bf
S}_{2j}$ and ${\bf S}_{2j+1}$ form a singlet pair
$(1/\sqrt{2})\,(\uparrow_{2j} \downarrow_{2j+1} - \downarrow_{2j}
\uparrow_{2j+1})$.
When $D=0$, therefore, Hamiltonian (1.1) smoothly connects the dimer
state of the spin-1/2 chain and the Haldane state of the spin-1 chain.

{}~~~~~Hida [3] numerically diagonalized the finite system of Hamiltonian
(1.1) in the $D=0$ case to find that there is no evidence for the phase
transition of the ground-sate between $J'=0$ and $J' = -\infty$.
Therefore he concluded that the Haldane phase of the spin-1 chain can be
interpreted as the special case of the dimer phase of the spin-1/2 chain.
This picture was supported by the work of Kohmoto and Tasaki [6], and
Takada [7] who used the nonlocal unitary transformation.
Hida [4] also performed the numerical diagonalization for the $D \ne 0$
case to draw the phase diagram of the $D$-model on the $J'-D$ plane.
He found the Haldane phase, the large-$D$ (LD) phase, the N\'eel I phase
and the N\'eel II phase.
The N\'eel I state is the usual N\'eel state of the spin-1/2 chain,
whereas the N\'eel II state is like the $|\cdots\uparrow_{2j-1}
\uparrow_{2j} \downarrow_{2j+1} \downarrow_{2j+2}\cdots\rangle$ state
which becomes the usual N\'eel state of the spin-1 chain in the limit of
$J' \rightarrow -\infty$.
Hida [8] also numerically estimated the critical exponents the
Haldane-LD boundary.

{}~~~~~Yamanaka, Hatsugai and Kohmoto [9] investigated the model
$$
  \eqalign{
    {\cal H}_\lambda
    &= 2J \sum_{j=1}^N (S_{2j}^x S_{2j+1}^x + S_{2j}^y S_{2j+1}^y
                       + \lambda S_{2j}^z S_{2j+1}^z )   \cr
    &~~~~~  + 2J' \sum_{j=1}^N {\bf S}_{2j-1} \cdot {\bf S}_{2j}~,}
    \eqno(1.3)
$$
where the parameter $\lambda$ represents the interaction anisotropy of
the antiferromagnetic bonds.
We call this model \lq\lq $\lambda$-model\rq\rq.
This model is equivalent to the spin-1 $XXZ$ model
$$
    \hat{\cal H}_\lambda^{S=1}
    = {J \over 2} \sum_{j=1}^N
      (\hat S_j^x \hat S_{j+1}^x + \hat S_j^y \hat S_{j+1}^y
       + \lambda \hat S_j^z \hat S_{j+1}^z)~,
    \eqno(1.4)
$$
 in the limit of $J' \rightarrow -\infty$.
We note that the ferromagnetic bonds should be isotropic so that ${\bf
S}_{2j}$ and ${\bf S}_{2j+1}$ form a triplet pair when $J' \rightarrow
-\infty$.
They mapped this model onto the highly anisotropic version of the
two-dimensional Ashkin-Teller model and performed the \lq\lq high
temperature expansion\rq\rq \ to obtain the ground-state phase diagram.
Their phase diagram consists of several phases, including the Haldane
phase, the N\'eel phase and the $XY$ phase.
Hida [5] also drew the phase diagram of the $\lambda$-model by mapping
the $\lambda$-model onto the $D$-model.
His conclusion was consistent with that of Yamanaka et al.[9].

{}~~~~~~Okamoto, Nishino and Saika [10] have shown that the Haldane-LD
boundary of Hida's phase diagram (obtained by the numerical calculations)
for the $D$-model can be semi-quantitatively reproduced by an analytical
method.
In this paper, we proceed their work to investigate the critical
properties of the Haldane-LD transition.
We show that the Haldane-LD transition is of the Gaussian universality
class and the critical exponents vary continuously on the boundary.
In \S 2, we explain the nature of the Haldane-LD transition.
In \S 3 the bosonization approach to the Haldane-LD transition given and
the critical properties are discussed.
The last section \S 4 is devoted to discussion.

\bigskip

{\bf\S 2. Nature of the Haldane-LD transition}
\medskip
{}~~~~~~In the following we investigate Hamiltonian (1.1) assuming
$J' < 0$.
Performing the spin rotation around the $z$-axis,
$$
  \matrix{
    & \tilde S_j^x = - S_j^x~,~~
    & \tilde S_j^y = - S_j^y~,~~
    & \tilde S_j^z =   S_j^z~,~~
    & {\rm for}~j=4l,4l+1~,  \cr
    & \tilde S_j^x =   S_j^x~,~~\hfill
    & \tilde S_j^y =   S_j^y~,~~\hfill
    & \tilde S_j^z =   S_j^z~,~~
    & {\rm otherwise}~,\hfill}
  \eqno(2.1)
$$
we can transform Hamiltonian (1.1) into an antiferromagnetic form [4]
$$
  \eqalign{
    \tilde{\cal H}
    = 2J_0 \sum_{j=1}^{2N}
           &\{ (\tilde S_j^x \tilde S_{j+1}^x
               + \tilde S_j^y \tilde S_{j+1}^y
               + \Delta \tilde S_j^z \tilde S_{j+1}^z)  \cr
           &~~~+ (-1)^j
               [\delta_\perp (\tilde S_j^x \tilde S_{j+1}^x
                              + \tilde S_j^y \tilde S_{j+1}^y)
                +\delta_z \tilde S_j^z \tilde S_{j+1}^z)]\}~, }
    \eqno(2.2)
$$
where
$$
  \eqalign{
    &J_0 = {J (1 + |\tilde J'|) \over 2}~,~~~~
      \Delta = 1 + {\tilde D - 2|\tilde J'| \over 1 + |\tilde J'|}~,
      \cr
    & \delta_\perp = {1 - |\tilde J'| \over 1 + |\tilde J'|}~,~~~~
     \delta_z = {1 - |\tilde J'| - (\tilde D - 2|\tilde J'|)
                 \over 1 + |\tilde J'|}~,}
  \eqno(2.3)
$$
and
$$
    \tilde J' \equiv J'/J~,~~~~~\tilde D \equiv D/J~.
    \eqno(2.4)
$$
Hamiltonian (2.3) can be interpreted as a generalized version of the
alternating antiferromagnetic Heisenberg chain with anisotropy (see
Fig.1).
The point $(|\tilde J'|,\tilde D)=(1,2)$ (called S point) is the
solvable point where Hamiltonian (2.2) represents the uniform and
isotropic antiferromagnetic chain.
The S point is the tricritical point of the Haldane phase, LD phase and
the N\'eel phase in Hida's phase diagram [4].

{}~~~~~In the bosonization theory of the bond alternation problem, it has
long been considered that the $\delta_z$ term plays an irrelevant role.
However, we have recently pointed out that the ground-state of
Hamiltonian (2.2) in the case of $\delta_\perp=0$ and $\delta_z>0$ is
the dimer state [11,12].
This shows the importance of the $\delta_z$ term, because the
ground-state would be the spin-fluid state if the $\delta_z$ term played
an irrelevant role.
When $\Delta=1$ and $\delta_\perp=\delta_z=1$, the ground-state of
Hamiltonian (2.3) is the complete dimer state (an array of local singlet
dimers), and both the $\delta_z$ term and the $\delta_\perp$ term give
the same contribution to the excitation gap apart from a factor 2.
Then, in $\Delta=1$ and $\delta_\perp = \delta_z > 0$ case where the
ground-state is the dimer state, we expect that the excitation gap
should behaves as
$$
   \epsilon_{\rm gap} \sim (2\delta_\perp + \delta_z)^{\nu}~,
   \eqno(2.5)
$$
due to the SU(2) symmetry.
Here $\nu$ is the exponent of the correlation length and we do not enter
into the logarithmic correction problem [13].
Thus we can consider that the $\delta_\perp$ term and the $\delta_z$
term are mutually cooperative when $\delta_\perp \delta_z > 0$, and
competing when $\delta_\perp \delta_z < 0$.
Of course, when $\Delta \ne 1$, equation (2.5) itself may be no longer
valid.

{}~~~~~Let us consider the $\Delta < 1$ case, where we can exclude the
possibility of the N\'eel state.
When $\delta_\perp > 0$ and $\delta_z > 0$, an effective singlet dimer
is formed by spins $\tilde{\bf S}_{2j}$ and $\tilde{\bf S}_{2j+1}$.
Their coupling is antiferromagnetic in the original spin (${\bf S}$)
representation before the transformation (2.1).
Since this singlet dimer is still a singlet dimer in the ${\bf S}$
representation, this state is the Haldane state.
On the other hand, when $\delta_\perp < 0$ and $\delta_z < 0$, spins
$\tilde{\bf S}_{2j-1}$ and $\tilde{\bf S}_{2j}$ (their coupling is
ferromagnetic in the ${\bf S}$ representation) form an effective singlet
pair.
By the transformation (2.1), this singlet dimer is transformed into the
triplet dimer (like $(1/\sqrt{2})(\uparrow_{2j-1} \downarrow_{2j} +
\downarrow_{2j-1} \uparrow_{2j})$ state) in the ${\bf S}$ representation.
This is the $\hat S^z=0$ state of the spin with $\hat S=1$.
Therefore this state is nothing but the LD state.
We note that, in the  $\tilde S$ representation, both the Haldane state
and the LD state are the singlet dimer state, but the dimer
configuration is different by one spin site between these two states.

{}~~~~~What happens in the region where $\delta_\perp<0$ and $\delta_z>0$ ?
If the effect of $\delta_z$ ($\delta_\perp$) is predominant, the
ground-state may be the Haldane (LD) state.
Therefore the ground-state phase transition between the Haldane state
and the LD state can be observed in this region.
The Haldane-LD phase boundary may be determined by the line on which the
effects of $\delta_\perp$ and $\delta_z$ cancel out with each other.
{}From equation (2.5), the simplest estimation for the Haldane-LD boundary
may be
$$
    2\delta_\perp + \delta_z = 0~,
    \eqno(2.6)
$$
which results in
$$
    \tilde D = 2|\tilde J'|~.
    \eqno(2.7)
$$
Although this estimation is too rough because equation (2.5) may be
valid only when $\Delta=1$, the nature of the Haldane-LD transition is
well explained by the above mentioned picture.
More elaborate estimation will be given in the next section.

\vfill\eject

{\bf\S 3. Critical properties of the Haldane-LD transition}
\medskip

{}~~~~~~Through a careful bosonization procedure, we can map Hamiltonian
(2.2) onto a generalized version of the sine-Gordon Hamiltonian
$$
    \tilde {\cal H}_{\rm b}
    = 2J_0 \int {\rm d} x
      \{ A(\nabla \theta)^2 + CP^2
         + B_{\rm I} \cos 2\theta - B_\perp \cos\theta
         - B_z (\nabla\theta)^2 \cos\theta\}~,
    \eqno(3.1)
$$
where the commutation relation
$$
    [\theta(x),P(x')] = {\rm i}\delta(x-x')~,
    \eqno(3.2)
$$
holds.
The coefficients $B_\perp,~B_z$ and $B_{\rm I}$ are obtained directly
from the bosonization procedure as
$$
    B_\perp = {\delta_\perp \over a}~,~~~~~
    B_z     = {\delta_z a \over \pi}~,~~~~~
    B_{\rm I}   = {\Delta \over 2a}~,
    \eqno(3.3)
$$
where $a$ is the spin spacing.
These expressions are considered to be valid when $\delta_\perp \ll 1,\
\delta_z \ll 1, \ \Delta \ll 1$.
Similar expressions were already obtained by Nakano and Fukuyama [14],
but there was an error in the sign in their expressions.
In their expressions, the $B_\perp$ term and the $B_z$ term are mutually
competing when $\delta_\perp \delta_z >0$.
{}From the discussion of \S 2, these terms should be mutually cooperative
when $\delta_\perp \delta_z >0$, as is realized in (3.1).

{}~~~~~~Since the term $B_{\rm I} \cos 2\theta$ is irrelevant for the
$\Delta<1$ case with which we are concerned, we may neglect this term by
setting
$$
    B_{\rm I} = 0~.
    \eqno(3.4)
$$
For the coefficients $A$ and $C$, the bosonization procedure leads to
$$
    A = {a \over 8\pi} \left(1 + {3\Delta \over \pi} \right)~,~~~~~
    C = 2\pi a \left( 1 - {\Delta \over \pi} \right)~.
    \eqno(3.5)
$$
Of course, these expressions may be valid for $\Delta \ll 1$.
Therefore we cannot use such expressions, because we require expressions
valid near $\Delta=1$ to discuss the Haldane-LD transition near the S
point ($\Delta =1, \ \delta_\perp = \delta_z = 0$).
When $\delta_\perp = \delta_z = 0$, Hamiltonian (2.2) represents the
uniform $XXZ$ chain for which the exact results are available.
In this case, the spin wave velocity $v$ [15], and the power decay
exponent $\eta$ [16], defined by $(-1)^r \ave{S_0^z S_r^z} \sim
r^{-\eta}$, are
$$
  \eqalignno{
    &{v \over 2J_0}
    = {\pi a \sqrt{1-\Delta^2} \over 2 \cos^{-1} \Delta}~,
    &(3.6)  \cr
    &\eta
    = {2 \over 1 + (\pi/2) \sin^{-1}\Delta}~,
    &(3.7)  }
$$
respectively.
If we use the bosonized Hamiltonian (3.1) with $B_{\rm I} = B_\perp =
B_z=0$, we obtain
$$
    {v \over 2J_0} = 2\sqrt{AC}~,~~~
    \eta = {1 \over 2\pi} \sqrt{C \over A}~.
    \eqno(3.8)
$$
{}From equations (3.6)-(3.8) we can immediately write down the expressions
for $A$ and $C$.
This procedure for the adjustment of the coefficients was first proposed
by Cross and Fisher [17] and applied by Nakano and Fukuyama [14,18].
If we expand $A$ and $C$ with respect to $\epsilon \equiv 1 - \Delta$,
we obtain
$$
  \eqalignno{
    &{v \over 2J_0} =  \,{\pi a \over 2} + O(\epsilon)~,~~~
     \eta = 1 + {\sqrt{2\epsilon} \over \pi}~,
    &(3.9)  \cr
    &A = {a \over 8} \left(1 - {\sqrt{2\epsilon} \over \pi} \right)~,~~~
    C = {\pi^2 a \over 2} \left(1 + {\sqrt{2\epsilon} \over \pi}
\right)~.    &(3.10)}
$$
up to the lowest order of $\epsilon$.
Expressions (3.9) and (3.10) were also obtained by Inagaki and Fukuyama
[19].
Thus our bosonized expression for the spin Hamiltonian (2.2) is equation
(3.1) with equations (3.3) and (3.10).
We note that this is valid for $|\delta_\perp| \ll 1,\ |\delta_z| \ll 1$
and $\epsilon \ll 1$.

{}~~~~~~To discuss the critical properties of the Haldane-LD transition,
we apply the self-consistent harmonic approximation (SCHA), which is
essentially the variational method, to the bosonized Hamiltonian (3.1)
when $\delta_\perp < 0$ and $\delta_z > 0$.
In the Haldane state, as is discussed in \S 2, the effect of $\delta_z$
is predominant and the average value of $\theta$ with respect to the
ground-state, $\ave{\theta}$, is
$$
    \ave{\theta} = 0~,~~~~~({\rm Haldane~state}),
    \eqno(3.11)
$$
so that the $B_z$ term gains the energy.
In the LD state, on the other hand, the effect of $\delta_z$ is
predominant and
$$
    \ave{\theta} = \pi~,~~~~~({\rm LD~state}).
    \eqno(3.12)
$$
Then we set the SCHA Hamiltonian
$$
    \tilde {\cal H}_{{\rm S}}
    = 2J_0 \int {\rm d} x
      \{ A(\nabla \phi)^2 + CP^2 + \tilde B \phi^2 \}~,
    \eqno(3.13)
$$
with
$$
    \phi \equiv
       \left \{ \matrix{ \theta~,     &\hbox{(Haldane state)}    \cr
                         \theta-\pi~, &\hbox{(LD state)} \hfill  }
       \right.~,
    \eqno(3.14)
$$
where $\tilde B$ is the variational parameter.
The parameter $\tilde B$ should be determined so that $\ave{\tilde {\cal
H}_{\rm b}}_{\rm S}$ is minimized, i.e.,
$$
    {\partial \ave{\tilde {\cal H}_{\rm b}}_{\rm S} \over \partial
\tilde B}
    = 0~,
    \eqno(3.15)
$$
where $\ave{\cdots}_{\rm S}$ denotes the average with respect to the
ground-state of $\tilde {\cal H}_{\rm S}$.

{}~~~~~~The excitation spectrum of $\tilde {\cal H}_{\rm S}$ is
$$
    \omega_{\rm S}(q)
    = v \sqrt{q^2 + q_{\rm c}^2}~,
    \eqno(3.16)
$$
where
$$
    q_{\rm c}^2 \equiv \tilde B/A~.
    \eqno(3.17)
$$
and $vq_{\rm c}$ is the excitation gap.
Since $\tilde {\cal H}_{\rm S}$ is harmonic, the relations
$$
  \eqalignno{
    &\ave{\exp({\rm i} u\phi)}_{\rm S}
    = \exp \left( -{u^2 \over 2} \ave{\phi^2}_{\rm S} \right)~,
    &(3.18)   \cr
    &\ave{(\nabla\phi)^2 \cos\phi}_{\rm S}
    = \ave{(\nabla\phi)^2}_{\rm S} \ave{\cos\phi}_{\rm S}~,
    &(3.19)   }
$$
hold with $u$ being a real number.
The average $\ave{(\nabla\theta)^2}_{\rm S}$ is
$$
  \eqalignno{
    &\ave{(\nabla\theta)^2}_{\rm S}
    = {C \over L} \sum_q {q^2 \over \omega_{\rm S}(q)}
    = {\eta \over 2}\,Q~,
    &(3.20)  \cr
    &Q
    \equiv \int_0^{\alpha_0^{-1}} {q^2 dq \over \sqrt{q^2 + q_{\rm c}^2}}
    \simeq {1 \over \alpha_0^2}~,~~~~
    (\alpha_0 q_{\rm c} \ll 1)~,
    &(3.21) }
$$
where $L$ is the system length, and the upper cutoff of the
$q$-summation is denoted by $\alpha_0^{-1}$ which may be proportional to
$a^{-1}$.
Luther and Peschel [20] suggested $\pi\alpha_0 = a$.
However, care must be taken to estimate $Q$, because equation (3.20) is
strongly dependent on $Q$.
Let us consider the $\Delta=1$ (i.e., $\eta=1$) case.
As discussed in \S 2, for $\Delta=1$, the effect of $\delta_\perp$ and
$\delta_z$ cancel out with each other when equation (2.6) holds.
Here we estimate so that the equation
$$
    \ave{B_\perp \cos\theta + B_z (\nabla\theta)^2 \cos\theta}_{\rm S}
    = 0~,
    \eqno(3.22)
$$
reproduces equation (2.6) when $\Delta = 1$, which yields
$$
    Q = \pi/a^2~.
    \eqno(3.23)
$$
After some calculations, for the Haldane state, we obtain
$$
  \eqalignno{
    {\partial \ave{\tilde {\cal H}_{\rm b}}_{\rm S} \over \partial
\tilde B}
    = &L \left\{   {B_\perp \over 2}
                   \exp\left( -{\ave{\phi^2}_{\rm S} \over 2} \right)
                 + {B_z \over 2}
                   \exp\left( -{\ave{\phi^2}_{\rm S} \over 2}  \right)
                   \ave{(\nabla\phi)^2}_{\rm S}
                 - \tilde B \right\}
      {\partial \ave{\phi^2}_{\rm S} \over \partial \tilde B}  \cr
      &- L B_z \exp\left( -{\ave{\phi^2}_{\rm S} \over 2} \right)
        {\partial \ave{(\nabla\phi)^2}_{\rm S} \over \partial \tilde B}~.
&(3.24)}
$$
The second term of the rhs of equation (3.24) can be dropped, because
$\ave{(\nabla\phi)^2}_{\rm S}$ is almost independent of $\tilde B$, as
can be seen equations (3.20) and (3.21).
Then, equation (3.15) with equation (3.24) is reduced to
$$
    \tilde B
    = {1 \over 2}\,\exp \left( -{\ave{\phi^2}_{\rm S} \over 2} \right)
      \left( B_\perp + {\pi\eta \over 2a^2}B_z \right)~.
    \eqno(3.25)
$$
The quantity $\ave{\phi^2}_{\rm S}$ is estimated as
$$
    \ave{\phi^2}_{\rm S}
    = {C \over L} \sum_q {1 \over \omega_{\rm S}(q)}
    = \eta \log{2\pi \over aq_{\rm c}}~,~~~~~(aq_{\rm c} \ll 1),
    \eqno(3.26)
$$
which leads to the self-consistent gap equation
$$
    Aq_{\rm c}^2
    = {1 \over 2a} \left(aq_{\rm c} \over 2\pi \right)^{\eta/2}\,
      \left(\delta_\perp + {\eta \over 2}\,\delta_z \right)~.
    \eqno(3.27)
$$

{}~~~~~~Similar calculation can be performed for the LD state.
If we replace $\delta_\perp + (\eta/2)\,\delta_z$ in equation (3.27) by
$|\delta_\perp + (\eta/2)\,\delta_z|$, the gap equation becomes valid
both for the Haldane state and for the LD state.
Thus the excitation gap behaves as
$$
    \epsilon_{\rm gap}
    =vq_{\rm c}
    \sim \left| \delta_\perp + {\eta \over 2}\, \delta_z \right|
         ^{2/(4-\eta)}~.
    \eqno(3.28)
$$
The Haldane-LD phase boundary can be obtained from $\epsilon_{\rm gap} =
0$, which leads to
$$
    2(1-|\tilde J'|)
    + (1 - \tilde D + |\tilde J'|)
      \left( 1 + {1 \over \pi}
                 \sqrt{2|\tilde J'| - \tilde D \over 1 + |\tilde J'|}
      \right)
    = 0~.
    \eqno(3.29)
$$
where equations (2.3), (3.3) and (3.9) are employed.
Okamoto, Nishino and Saika [10] have already obtained this phase
boundary equation, compared it with Hida's numerical result [8] and
discussed its validity.

{}~~~~~If the critical value of $D$ is denoted by $D_{\rm c}$ when $|J'|$
is fixed, we can rewrite equation (3.28) as
$$
  \eqalignno{
    &\epsilon_{\rm gap}
    \sim |D-D_{\rm c}|^\nu~,
    &(3.30)  \cr
    &\nu
    = {2 \over 4-\eta}~.
    &(3.31)  }
$$
Thus we obtain
$$
    \nu_{\rm H}
    = \nu_{\rm LD}
    = \nu~,
    \eqno(3.32)
$$
where $\nu_{\rm H}$ ($\nu_{\rm LD}$) is the critical exponent of the
excitation gap when the Haldane-LD boundary is approached from the
Haldane (LD) phase.

{}~~~~~~The calculation of the longitudinal spin correlation $\ave{S_j^z
S_l^z}$ by the use of $\tilde {\cal H}_{\rm S}$ has been already done by
the present author [21].
Here we only write down the final expression for $\ave{S_j^z S_l^z}$,
without entering into details;
$$
    \ave{S_j^z S_l^z}
    \sim {(-1)^{|j-l|} \over \sqrt{|j-l|}}
         \exp \left( -{|j-l|a \over \xi} \right)~,~~~~~
    (|j-l| \rightarrow \infty)~,
    \eqno(3.33)
$$
where $\xi$ is the correlation length defined by
$$
    \xi
    = {v \over \epsilon_{\rm gap}}
    \sim \left| \delta_\perp + {\eta \over 2} \,\delta_z \right|
         ^{-\nu}
    \sim |D - D_{\rm c}|^{-\nu}~.
    \eqno(3.34)
$$
Thus $\nu$ is also the exponent of the correlation length as expected.
Comparing the result of reference [21] with the exact result for the
$XY$ case [22], we may rely on the exponential factor in equation (3.33),
although the preceding power factor is not reliable.

{}~~~~~~Hida [8] proposed the string order parameters for the Haldane
phase and the LD phase.
In the ${\tilde S}$ representation, these string order parameters are
defined by
$$
  \eqalignno{
    O_{\rm H}^\alpha
    &= \lim_{|j-l| \rightarrow \infty} O_{\rm H}^\alpha(j-l)~,
    &(3.35)   \cr
    O_{{\rm L}{\rm D}}^\alpha
    &= \lim_{|j-l| \rightarrow \infty} O_{{\rm L}{\rm D}}^\alpha(j-l)~,
    &(3.36)   }
$$
with
$$
  \eqalignno{
    O_{\rm H}^\alpha(j-l)
    &= -4 \ave{{\tilde S}_{2j}^\alpha
              \exp\{{\rm i}\pi({\tilde S}_{2j+1}^\alpha + {\tilde
S}_{2j+2}^\alpha
                           + \cdots + {\tilde S}_{2j-2}^\alpha) \}
              {\tilde S}_{2l-1}^\alpha}~,~~~~~
    &(3.37)  \cr
    O_{{\rm L}{\rm D}}^\alpha(j-l)
    &= -4 \ave{{\tilde S}_{2j-1}^\alpha
              \exp\{{\rm i}\pi({\tilde S}_{2j}^\alpha + {\tilde
S}_{2j+1}^\alpha
                           + \cdots + {\tilde S}_{2l-1}^\alpha) \}
              {\tilde S}_{2l}^\alpha}~,~~~~~
    &(3.38)  \cr
    \alpha
    &= x,y,z~.
    &(3.39)  }
$$
By the use of the identity $S_j^z = \exp({\rm i}{\tilde S}_j^z)/2{\rm
i}$ valid for the spin-1/2 operators, we can rewrite equations (3.37)
and (3.38) as
$$
 \eqalignno{
    O_{\rm H}^z(j-l)
    &= \ave{ \exp\{{\rm i}\pi({\tilde S}_{2j}^z + {\tilde S}_{2j+1}^z
                           + \cdots + {\tilde S}_{2l-1}^z) \}}~,
    &(3.40)  \cr
    O_{{\rm L}{\rm D}}^z(j-l)
    &= \ave{ \exp\{{\rm i}\pi({\tilde S}_{2j-1}^z + {\tilde S}_{2j}^z
                           + \cdots + {\tilde S}_{2l}^z) \}}~,
    &(3.41) }
$$
respectively.
Since the slowly varying part of the $z$-component of the spin density
is expressed as $(1/2\pi)\,(\partial\theta/\partial x)$,
the boson representations of $O_{\rm H}^z(j-l)$ and $O_{{\rm L}{\rm
D}}^z(j-l)$ are
$$
  \eqalignno{
    O_{\rm H}^z(x-x')
    &= O_{{\rm L}{\rm D}}^z(x-x')  \cr
    &= \ave{ \exp \{ {\rm i}[\theta(x) - \theta(x')]/2 \}}  \cr
    &= \ave{ \exp \{ {\rm i}[\phi(x) - \phi(x')]/2 \}}      \cr
    &\equiv O^z(x-x')~.
    &(3.42) }
$$
We note that the difference between the $O_{\rm H}^z(x-x')$ and $O_{{\rm
L}{\rm D}}^z(x-x')$ is lost due to the continuum approximation used in
the bosonization method.

{}~~~~~~Let us calculate $O^z$ in the framework of the SCHA.
Due to the harmonic nature of $\tilde {\cal H}_{\rm S}$, we have
$$
  \eqalignno{
    O^z(x-x')
    &= \exp \left\{ -{1 \over 8}
                   \ave{[ \theta(x)-\theta(x')]^2 }_{\rm S}
           \right\}  \cr
    &= \exp \left( -{1 \over 4} \ave{\phi^2}_{\rm S} \right)\,
      \exp \left\{ {1 \over 4} \ave{\phi(x)\phi(x')}_{\rm S} \right\}~.
    &(3.43) }
$$
The average $\ave{\phi(x)\phi(x')}_{\rm S}$ is estimated as [21]
$$
  \eqalignno{
    &\ave{\phi(x)\phi(x')}_{\rm S}
    = {C \over \pi v}
       K_0(q_{\rm c}[|x-x'|+\alpha_1])~,
    &(3.44)  \cr
    &\alpha_1
    = {a {\rm e}^{-\gamma} \over \pi}~,~~~~~
    \gamma = 0.5772\cdots~~\hbox{(Euler's constant)}~,
    &(3.45)  }
$$
where $K_n(y)$ is the $n$-th order modified Bessel function of the
second kind.
We note that the parameter $\alpha_1$ is determined so that equation
(3.26) is reproduced when $|x-x'|=0$.
Using the asymptotic behavior of $K_0(y)$
$$
    K_0(y)
    \simeq \sqrt{\pi \over 2y} {\rm e}^{-y}~,~~~~~
    (y \rightarrow \infty)~,
    \eqno(3.46)
$$
we obtain
$$
  \eqalignno{
    O^z
    &= \exp \left( -{1 \over 4} \ave{\phi^2}_{\rm S} \right)  \cr
    &\sim q_{\rm c}^{\eta/4}
    \sim |D-D_{\rm c}|^{\eta/(8-2\eta)}
    &(3.47)  }
$$
If we denote the exponents of $O_{\rm H}^z$ and $O_{{\rm L}{\rm D}}^z$
by $2\beta^z_{\rm H}$ and $2\beta^z_{{\rm L}{\rm D}}$ respectively (for
the factor 2, we follow Hida's definition [8], we see
$$
    \beta_{\rm H}
    = \beta_{{\rm L}{\rm D}}
    = {\eta \over 16-4\eta}
    \equiv \beta~.
    \eqno(3.48)
$$

{}~~~~~~We can also calculate the asymptotic behavior of $O^z(x-x')$ when
$|x-x'| \rightarrow \infty$ just on the Haldane-LD boundary.
Using equations (3.43)-(3.45) and taking the limit $q_{\rm c}
\rightarrow 0$, we obtain
$$
    O^z(x-x')|_{D=D_{\rm c}}
    \sim |x-x'|^{-\eta/4}~,
    \eqno(3.49)
$$
where the asymptotic form
$$
    K_0(y) \simeq -\log y~,~~~~~
    (y \rightarrow 0)~,
    \eqno(3.50)
$$
is employed.
Hida [8] also numerically calculated the system size dependence of
$O_{\rm H}^z$ and $O_{{\rm L}{\rm D}}^z$ on the Haldane-LD boundary.
He defined
$$
  \eqalignno{
    &O_{\rm H}^z(N)|_{D=D_{\rm c}}
    \sim N^{-\mu_{\rm H}}~,
    &(3.51)  \cr
    &O_{{\rm L}{\rm D}}^z(N)|_{D=D_{\rm c}}
    \sim N^{-\mu_{{\rm L}{\rm D}}}~,
    &(3.52)  }
$$
where $N$ is the system size.
Although Hida [8] used $\eta^z$ for this exponents, we use $\mu$ to
avoid confusion.
{}From equation (3.49), we can expect
$$
    \mu_{\rm H}
    = \mu_{{\rm L}{\rm D}}
    = {\eta \over 4}
    \equiv \mu~.
    \eqno(3.53)
$$
This result can also obtained from the finite size scaling ansatz
$$
    O^z(|D-D_{\rm c}|,N)
    \sim N^{-\mu} f(N|D-D_{\rm c}|^\nu)~,
    \eqno(3.54)
$$
where we note that $N/\xi \sim N|D-D_{\rm c}|^\nu$.
The form of equation (3.54) is chosen so that equations (3.51) and (3.52)
are reproduced at $D=D_{\rm c}$.
If we fix $|D-D_{\rm c}|$ and take the limit $N \rightarrow \infty$, we
obtain
$$
  \eqalignno{
    &f(x) \sim x^\mu~,~~~~~~(x \rightarrow \infty)~,
    &(3.55)  \cr
    &O^z(|D-D_{\rm c}|,N=\infty)
    \sim |D-D_{\rm c}|^{\mu\nu}~,
    &(3.56)  }
$$
from the condition that $O^z(|D-D_{\rm c}|,N)$ should be independent of
$N$ when $N \rightarrow \infty$.
{}From equations (3.47) and (3.48) it follows that
$$
    \mu\nu = 2\beta~,
    \eqno(3.57)
$$
which readily leads to equation (3.53).
The relation (3.57) was also noticed by Hida [8].

\bigskip

{\bf\S 4. Discussion}
\medskip

{}~~~~~~We have calculated several critical exponents of the Haldane-LD
transition.
These exponents are controlled by $\nu$, which varies continuously on
the Haldane-LD boundary.
Thus the present transition is of the Gaussian universality class.
This fact has been already pointed out by several groups [8-10].

{}~~~~~~In this paper we have used the mapping of the original spin
Hamiltonian onto the generalized version of the sine-Gordon Hamiltonian.
Since several approximation have been used in the course of the mapping,
the values of $D_{\rm c}$ and the critical exponent themselves are
unreliable.
We believe, however, that the present theory well describes the
qualitative feature of the Haldane-LD transition.
{}From equations (3.31), (3.48) and (3.53), our prediction for the
relations between critical exponents are
$$
  \eqalignno{
    &\nu_{\rm H}
    = \nu_{{\rm L}{\rm D}}
    = \nu~,
    &(4.1)  \cr
    &\beta_{\rm H}
    = \beta_{{\rm L}{\rm D}}
    = \beta
    = {2\nu-1 \over 4}~,
    &(4.2)  \cr
    &\mu_{\rm H}
    = \mu_{{\rm L}{\rm D}}
    = \mu
    = 1 - {1 \over 2\nu}~.
    &(4.3)  }
$$
The reliability of the present theory can be examined by testing whether
equations (4.2) and (4.3) hold or not in Hida's numerical result [8,23].
This test is summarized in Table I, where the exponents $\nu,\
\beta_{\rm H},\allowbreak \ \beta_{\rm LD},\ \mu_{\rm H}$ and $\mu_{\rm
LD}$ are numerical results by Hida and $\beta_{\rm theor}$ and $\mu_{\rm
theor}$ are calculated from Hida's $\nu$ through equations (4.2) and
(4.3).
As can be seen, the values of $\beta_{\rm theor}$ and $\mu_{\rm theor}$
very well agree with $\beta_{\rm H}$ and $\beta_{{\rm L}{\rm D}}$, and
$\mu_{\rm H}$ and $\mu_{{\rm L}{\rm D}}$, respectively.
Therefore we can say that our theory successfully describes the
qualitative feature of the Haldane-LD transition.

\bigskip

{\bf Acknowledgements}

{}~~~~~~I would like to express my appreciation to K. Hida for useful
discussions and providing me his numerical data.
I also thank K. Nomura, Y. Saika and D. Nishino for stimulating
discussions.

\vfill\eject

\parindent=30truept
\noindent
{\bf References}

\medskip

\item{[1]} Haldane F D M 1983 {\it Phys. Lett}. {\bf 93A} 464.

\item{[2]} Haldane F D M 1983 {\it Phys. Rev. Lett.} {\bf 50} 1153.

\item{[3]} Hida K 1992 {\it Phys. Rev. B} {\bf 45} 2207.

\item{[4]} Hida K 1992 {\it Phys. Rev. B} {\bf 46} 8268.

\item{[5]} Hida K 1993 {\it J. Phys. Soc. Jpn.} {\bf 62} 1463.

\item{[6]} Kohmoto M and Tasaki H 1992 {\it Phys. Rev. B} {\bf 46} 3486.

\item{[7]} Takada S 1992 {\it J. Phys. Soc. Jpn.} {\bf 61} 428.

\item{[8]} Hida K 1993 {\it J. Phys. Soc. Jpn.} {\bf 62} 1466.

\item{[9]} Yamanaka M,  Hatsugai Y and Kohmoto M 1993 {\it Phys. Rev. B}
{\bf 48} 9555.

\item{[10]} Okamoto K, Nishino D and  Saika Y 1993 {\it J. Phys. Soc.
Jpn.} {\bf 62} 2587.

\item{[11]} Okamoto K 1992 {\it J. Phys. Soc. Jpn.} {\bf 61} 3488.

\item{[12]} Okamoto K and Nishino D 1993 {\it Solid State Commun.} {\bf
85} 343.

\item{[13]} Black J L and Emery V J 1981 {\it Phys. Rev. B} {\bf 23} 429.

\item{[14]} Nakano T and Fukuyama H 1981 {\it J. Phys. Soc. Jpn.} {\bf
50} 2489.

\item{[15]} des Cloizeaux J and Gaudin M 1966 {\it J. Math. Phys.} {\bf
7} 1384.

\item{[16]} Johnson J D, Krinsky S and McCoy B 1973 {\it Phys. Rev A}
{\bf 8} 2526.

\item{[17]} Cross M C and Fisher D S 1979 {\it Phys. Rev. B} {\bf 19}
402.

\item{[18]} Nakano T and Fukuyama H 1981 {\it J. Phys. Soc. Jpn.} {\bf
49} 1679.

\item{[19]} Inagaki S and Fukuyama H 1983 {\it J. Phys. Soc. Jpn.} {\bf
52} 2504.

\item{[20]} Luther A and  Peschel I 1975 {\it Phys. Rev. B} {\bf 12}
3908.

\item{[21]} Okamoto K 1987 {\it J. Phys. Soc. Jpn.} {\bf 56} 912.

\item{[22]} Okamoto K 1988 {\it J. Phys. Soc. Jpn.} {\bf 57} 1988.

\item{[23]} Hida K: private communications.

\vfill\eject

\parindent=45truept
\noindent
Figure Captions:

\bigskip
\item{Fig.1} Important parameters in Hamiltonian (2.2) on the $|\tilde
J'|-\tilde D$ plane.
The transverse and longitudinal alternations are mutually competing in
the shadowed areas.
The point ${\rm S}\,(|\tilde J'|=1,\tilde D=2)$ corresponds to the
uniform isotropic antiferromagnetic Heisenberg model.
\end